\documentclass[ajl]{emulateapj}

\usepackage{graphicx,epsfig,natbib,color}
\usepackage{lscape}
\usepackage{graphicx}
\usepackage{epsfig}
\usepackage{natbib}

\usepackage[usenames,dvipsnames,svgnames,table]{xcolor}
\usepackage[colorlinks=true,
           linkcolor=blue,
           urlcolor=blue,
           citecolor=blue]{hyperref}

\newcommand{\kms}{km~s$^{-1}$}

\begin{document}

\title{SDO/AIA OBSERVATIONS OF A REFLECTING LONGITUDINAL WAVE IN A CORONAL LOOP}
\author{PANKAJ KUMAR\altaffilmark{1,2}, D.E. INNES\altaffilmark{2}, B. INHESTER\altaffilmark{2}}
\affil{$^1$Korea Astronomy and Space Science Institute (KASI), Daejeon, 305-348, Republic of Korea}
\affil{$^2$Max-Planck Institut f\"{u}r Sonnensystemforschung, 37191 Katlenburg-Lindau, Germany}
\email{pankaj@kasi.re.kr}

\begin{abstract}
We report high resolution observations from the {\it Solar Dynamics Observatory/Atmospheric Imaging Assembly} (AIA) of intensity oscillations in a  hot, $T \sim$8-10~MK, loop. The AIA images show a large coronal loop that was rapidly heated following plasma ejection from one of the loop's footpoints.  A wave-like intensity enhancement, seen very clearly in the 131 and 94~\AA\ channel images, propagated ahead of the ejecta along the loop, and was reflected at the opposite footpoint. The wave reflected four times before fading. It was only seen in the hot, 131 and 94~\AA\ channels.
 The characteristic period and the decay time of the oscillation was  $\sim$630 and $\sim$440 s, respectively.  The phase speed was about 460--510~\kms\ which roughly matches the sound speed of the loop (430-480~\kms).
The observed properties of the oscillation are consistent with the observations of Doppler shift oscillations discovered by the {\it Solar and Heliospheric Observatory/Solar Ultraviolet Measurement of Emitted Radiation} (SUMER) and with their interpretation as slow magnetoacoustic waves. We suggest that the impulsive injection of plasma, following reconnection at one of the loop footpoints, led to rapid heating and the
propagation of a longitudinal compressive wave along the loop. The wave bounces back and forth a couple of times before fading.

\end{abstract}

\section{INTRODUCTION}
 MHD waves and oscillations observed in the solar corona are extremely important as they provide an excellent opportunity to probe the corona indirectly via coronal seismology \citep{roberts1984,nakariakov2005,demoortel2009}.

Slow magnetoacoustic waves were discovered in post-flare coronal loops with SOHO/SUMER, by measuring periodic Doppler shifts in lines from  Fe XIX and Fe XXI, formed at temperatures, $T>6$~MK \citep{kliem2002,wang2002,wang2003a,wang2003b}.  A survey of over 50 events
found  oscillation periods with $\sim$7-31 minutes, decay times of $\sim$6-37 minutes, and maximum Doppler velocities in the range 100--300~km~s$^{-1}$ \citep{wang2005}.
These oscillations were interpreted as standing slow magnetoacoustic modes because their phase speed was close to sound speed in the loop, and in one event there was a quarter period phase shift between the observed velocity and intensity oscillations. Although indicative of a standing mode, such a phase shift could also be produced by loops moving into and out of a spatial pixel as a result of Alfv\'enic oscillations \citep{tian2012}.
Similar Doppler shift oscillations were also observed in flare and coronal emission lines with Yohkoh/BCS and Hinode/EIS, respectively \citep{mariska2005,mariska2006,mariska2008}. Recently, \citet{kim2012} reported observations of slow magnetoacoustic oscillations in 17 GHz Nobeyama Radioheliograph  density and AIA 335~\AA\ measurements, during an M1.6 flare.

It was suggested that the waves are excited by small flares at one of the footpoints of the heated loop \citep{wang2003a,wang2005,wang2011} because a RHESSI hard X-ray source was often seen near one of the loops' footpoints. In addition many events showed that there were initially two spectral components suggesting that the wave onset was accompanied by a pulse of hot plasma.

Numerical magnetohydrodynamic (MHD) simulations demonstrated that  standing slow-mode magnetoacoustic waves can be excited by a localized pressure pulse at one of the footpoints of a loop \citep{selwa2005, selwa2007, taroyan2005}.
\citet{ofman2012} recently performed three-dimensional MHD modelling of a bipolar active region and concluded that the excitation of slow-mode (and some transverse) oscillations in coronal loops may result from the injection of plasma at the corona--chromosphere interface of the loop footpoints.

The strong damping of Doppler-shift oscillations was investigated by \citet{ofman2002}. The authors suggested that thermal conduction is the main dissipation mechanism for the slow magnetoacoustic waves in hot loops. Later numerical studies showed that  other  physical effects, (e.g., viscosity, radiative emission,
 shock dissipation, coupling between fast- and slow-mode MHD waves, wave leakage etc)  also play a role \citep{briceno2004,pandey2006,bradshaw2008,haynes2008,verwichte2008,
 Ogrodowczyk2007,Ogrodowczyk2009,selwa2009}.
After all, the mechanisms  causing the observed very rapid excitation and damping of standing slow magnetoacoustic waves in hot loops are still not well understood.

There is considerable support for the slow magnetoacoustic wave interpretation of the Doppler-shift
oscillations seen in hot coronal loops. One of the most critical and least constrained observation parameter for this interpretation is the loop length. Many of the derived speeds were close to the sound speed and if the loops were actually 20\%\ longer than assumed, the derived wave speeds would be supersonic. Most Doppler-shift oscillations were seen in loops on the solar limb where there are large uncertainties in the loop length estimation.

The loop oscillation, reported here, occurred in  an active region  50$^\circ$ east of central meridian, and we are able to clearly identify the loop along its entire length, including its two footpoints.  The oscillation period and the loop length are  well constrained and we find that the phase speed of the oscillation was close to, or slightly faster, than the loop's
sound speed.

In this letter, we report the trigger and observation of a reflecting longitudinal wave, seen with AIA, in a hot loop observed on 7 May 2012. The event occurred shortly after and close to the site of a C-class flare. In section 2, we present the observations and results and in the last section, we discuss and summarize the results.

\section{OBSERVATIONS AND RESULTS}
The {\it Atmospheric Image Assembly} (AIA; \citealt{lemen2012}) onboard the {\it Solar Dynamics Observatory} (SDO; \citealt{pesnell2012})  
obtains full disk images of the Sun (field-of-view $\sim$1.3 R$_\odot$) with a spatial resolution of 1.5$\arcsec$ (0.6$\arcsec$  pixel$^{-1}$) 
and a cadence of 12 sec in 10 extreme ultraviolet (EUV) and UV filters. For the present study, we utilized 171~\AA\ (Fe IX, with formation temperature $T\approx$0.7 MK), 131~\AA\ (Fe VIII/XXI, $T\approx$0.4 \& 11 MK), and 304~\AA\ (He II, $T\approx$0.05 MK) images. We also used Helioseismic and Magnetic Imager (HMI; \citealt{scherrer2012}) magnetograms to explore the magnetic field configuration of the active region (AR).

\begin{figure*}
\centering{
\includegraphics[width=8.5cm]{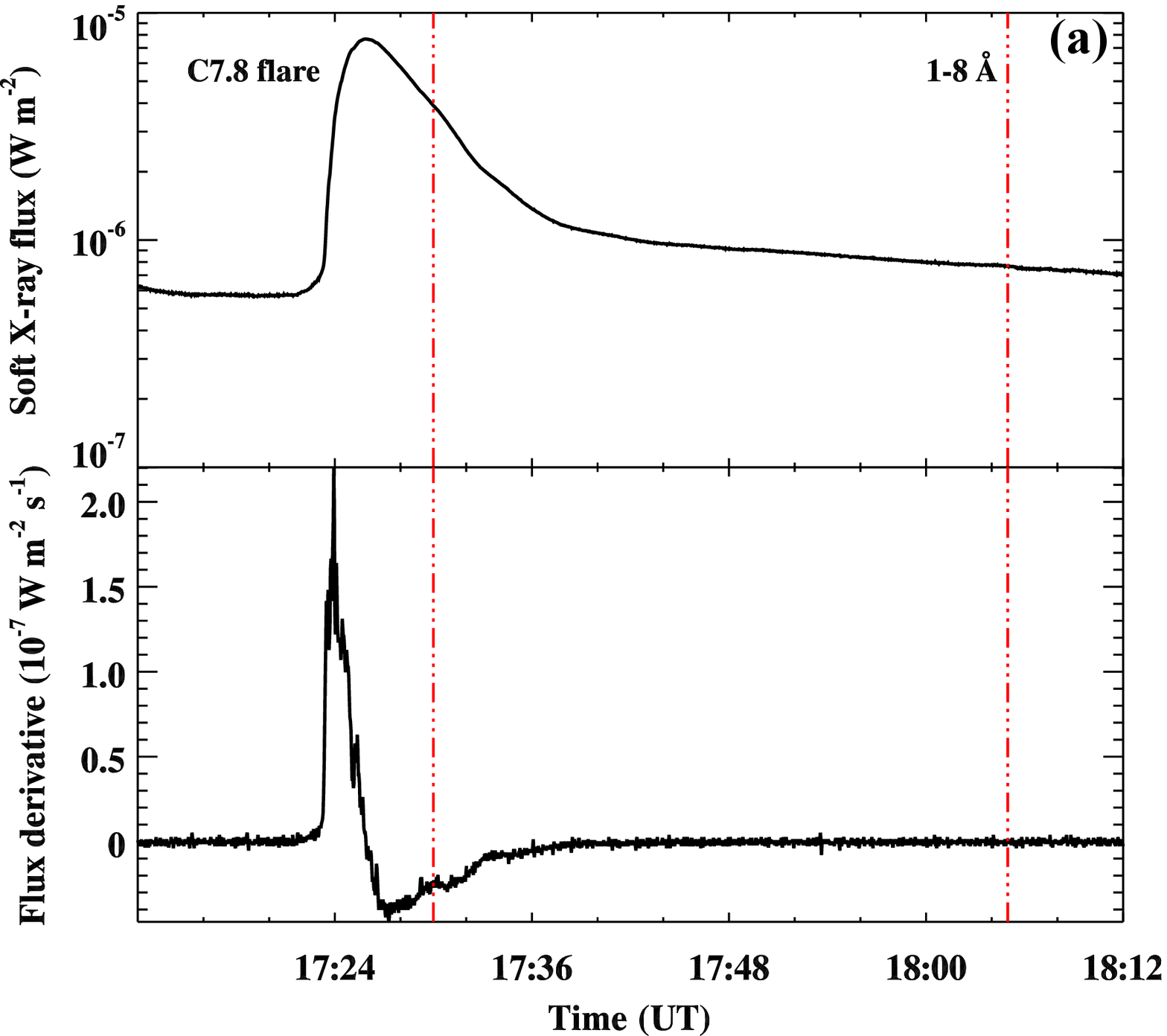}
\includegraphics[width=9cm]{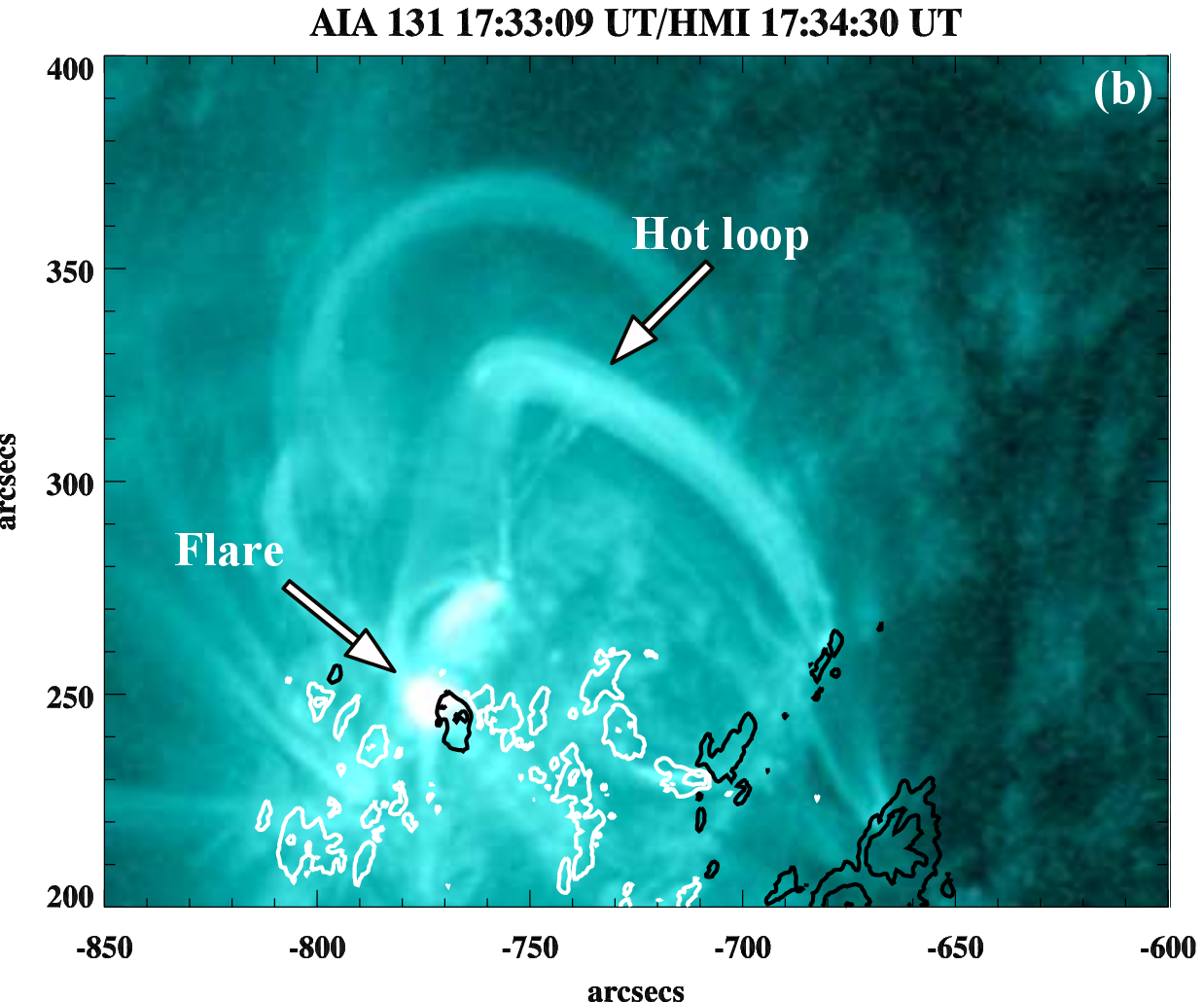}
}
\caption{(a) GOES Soft X-ray flux profile in 1-8 \AA~ channel at the time of the C7.8 flare (top panel) and soft X-ray flux-derivative (bottom panel). Two vertical dotted lines (red) represent the time interval of the observed hot EUV loop (in 131 \AA). (b) AIA 131 \AA~ image of the flare site and associated hot loop system. This image is overlaid by the HMI magnetogram contours of positive (white) and negative (black) polarities. The contour levels are $\pm$400, $\pm$800, and $\pm$1200 G.}
\label{fig1}
\end{figure*}

The AR NOAA 11476 produced an impulsive C7.4 class flare which started at $\sim$17:20~UT, peaked at $\sim$17:26~UT, and ended at $\sim$17:36~UT. The top panel of Figure \ref{fig1}(a) displays the GOES soft X-ray flux profile in 1-8~\AA\ channel. The bottom panel shows the soft X-ray flux derivative, which may be considered as a proxy of the hard X-ray burst during the flare impulsive phase \citep{neupert1968}.

The loop intensity oscillation was detected 6~min after the flare peak. Two vertical dotted red lines represent the time over which the loop oscillation was detectable.
Figure \ref{fig1}(b) shows the loop in  the AIA 131~\AA\ image at 17:33:09~UT. The flare was triggered at the eastern footpoint of the hot coronal loop (marked by an arrow). To view the magnetic field distribution at the flare site, we overlaid HMI magnetogram contours of positive (white) and negative (black) polarities on the AIA image. The presence of opposite polarity (negative) field is evident at the flare site.

\subsection{Oscillation characteristics}

The plasma ejection, loop heating and wave  appeared shortly after the flare at $\sim$17:30~UT. This loop was detected only in the AIA 131 and 94~\AA\ channels, indicating that it had a  high temperature.

\begin{figure*}
\centering{
\includegraphics[width=12cm]{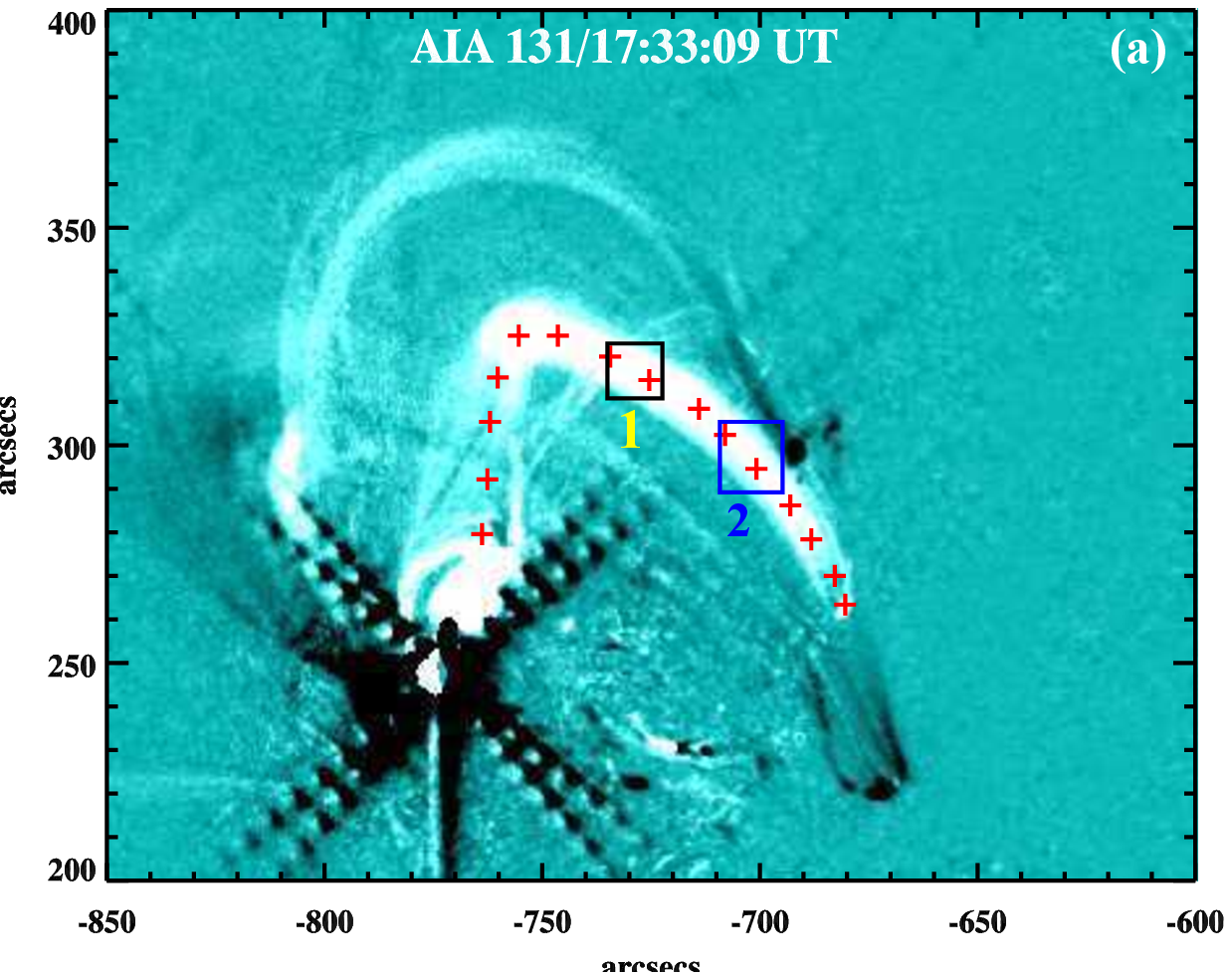}

\includegraphics[width=6.5cm]{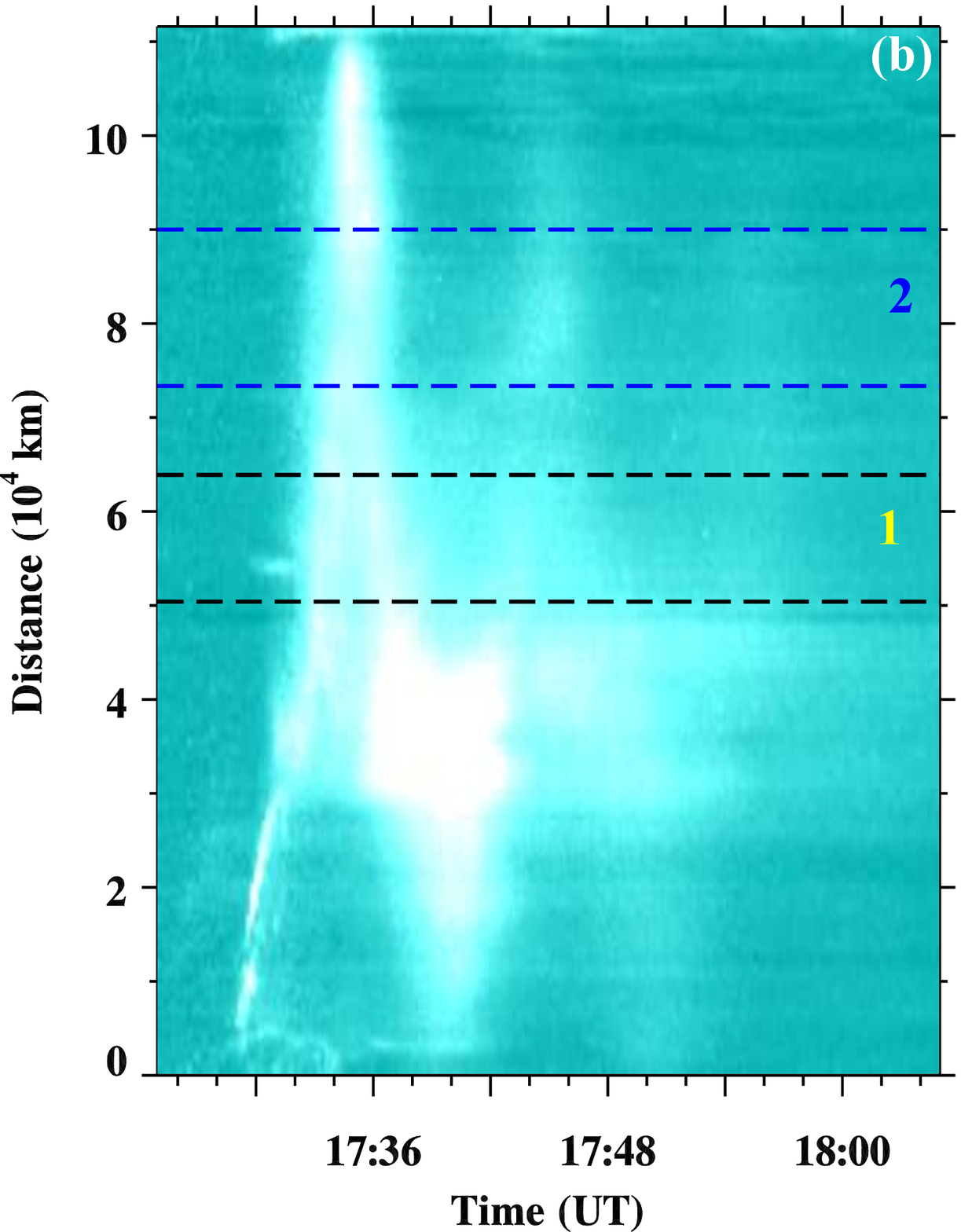}
\includegraphics[width=6.9cm]{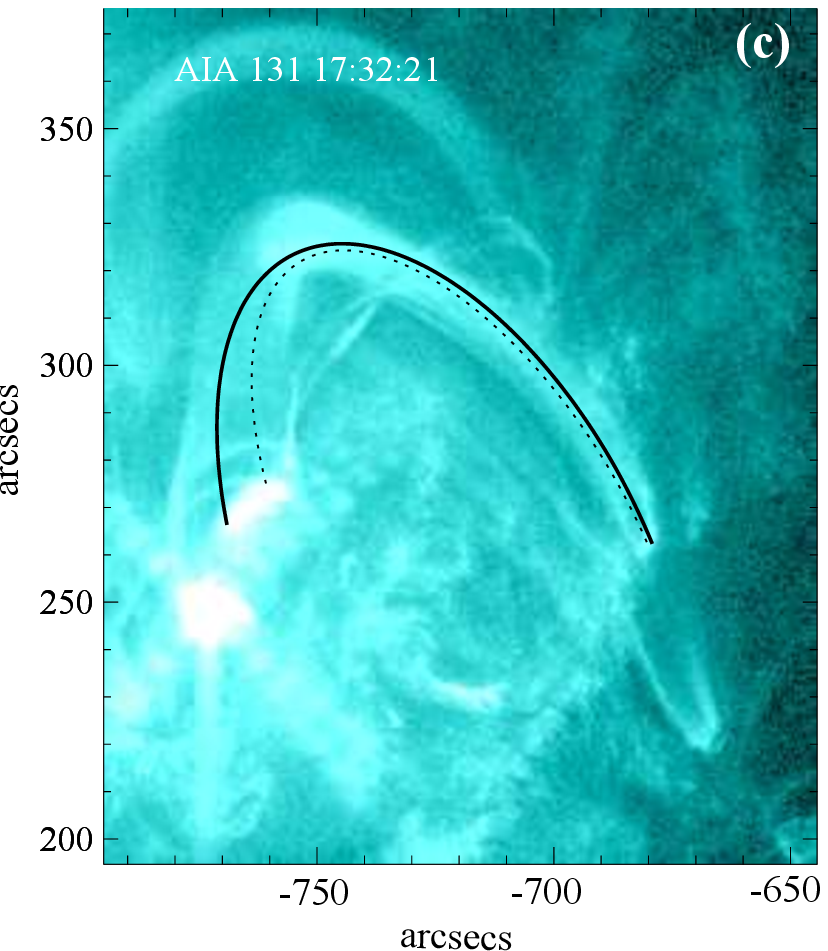}
}
\caption{(a) AIA 131 \AA~ base difference image of the hot loop. The red `+' symbols display the selected path along the loop used for the stack plot. (b) Time-distance plot showing the intensity evolution along the selected path of the hot loop. (c) The loop overlaid with the best-fit (black solid line) and shortest (dotted line) loop fits.}
\label{fig2}
\end{figure*}

Figure~\ref{fig2}(a) shows the AIA 131~\AA\ base-difference image of the loop just after the first loop crossing by the wave. To investigate the oscillations, we chose the path along the loop  marked by red `+' symbols and extracted the 131~\AA\ base-difference intensity  between 17:25--18:05~UT. The  base image was taken at 17:24:45~UT.
 The resulting time-distance plot of the intensity distribution is shown in Figure \ref{fig2}(b). This plot clearly reveals an intensity oscillation along the loop length. These oscillations are also clearly visible in the 131 and 94~\AA\ movie, available online. The hot-plasma emission started along the eastern leg, close to where the C-class flare was triggered, and then propagated along the loop to the other footpoint where it was reflected back along the loop. Decaying, multiple reflections continued for about 30~min, until about 18:00~UT. In total, the intensity oscillations went to and fro two and a half times.

\begin{figure*}
\centering{
\includegraphics[width=14cm]{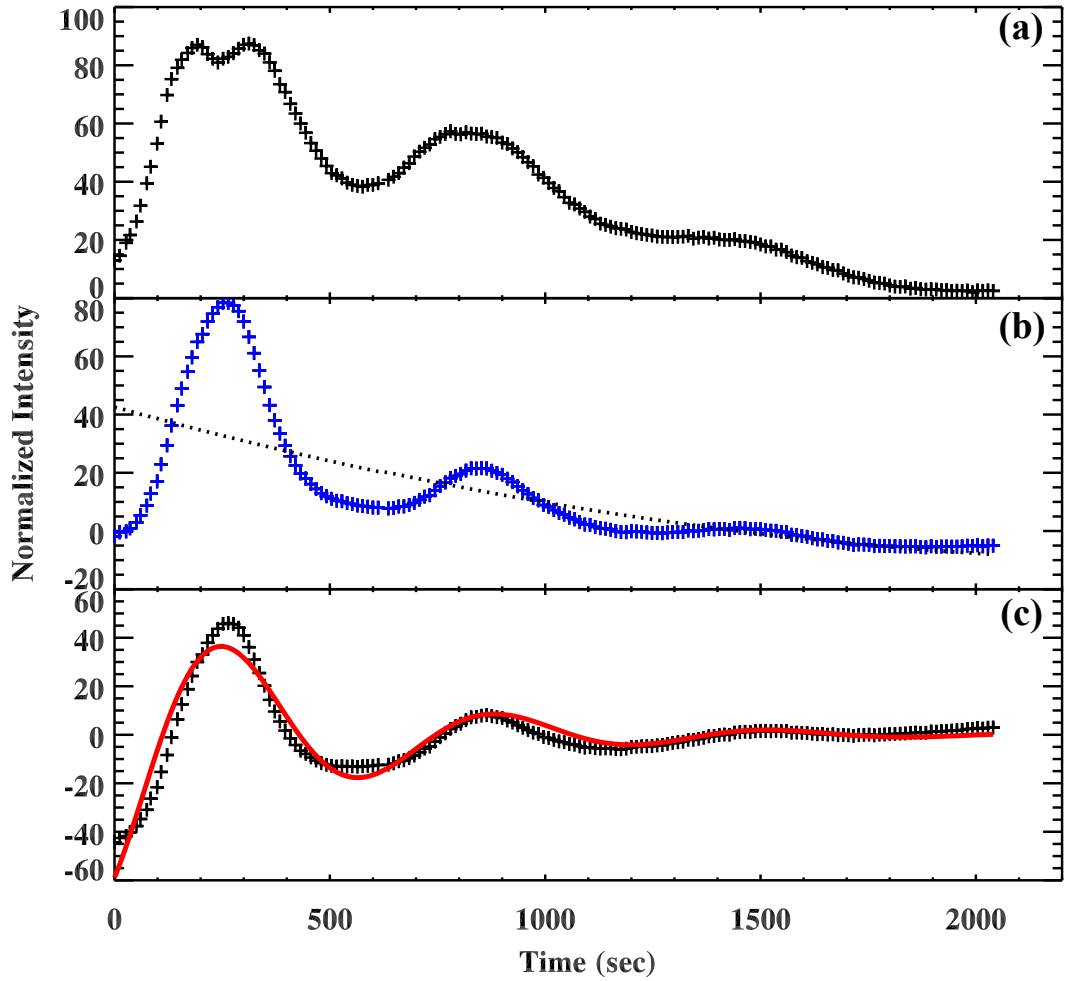}
}
\caption{(a-b) Time profiles of the average counts extracted from the box regions 1 and 2 in Figure \ref{fig2}a. The dotted line is the second order parabolic trend.
(c) Detrended light curve after removing the second order parabolic trend. The thick red curve shows a best fit on it.
The start time of the profiles is 17:30:57 UT.}
\label{fig3}
\end{figure*}

To study the oscillation properties, we extracted the mean intensity within the boxes 1 and 2
(shown in Figure~\ref{fig2}) from the 131~\AA\ base-difference images. Figure~\ref{fig3}(a) and (b) display the normalized intensity profiles within the boxes 1 and 2, respectively. The top panel (a) exhibits double or broader peaks because the incident and reflected waves are partially resolved  at  this  position (see Figure~\ref{fig2}). Box 2 is close to the western footpoint so the incident and reflected wave brightening merges to a single peak. At both positions, the intensity oscillation decayed rapidly.  For the purpose of fitting  and to emphasize the oscillations, we de-trended the intensity curve by subtracting a parabola (marked by the blue dotted line in panel (b)).
We then fitted the de-trended light curve, $I(t)$,  with the function

\begin{equation}
I(t)=A\mbox{ sin}(\frac{2\pi t }{P}+\phi)\mbox{ exp}(\frac{-t}{\tau}),
\end{equation}

\noindent
where $A$, $P$, $\tau$, and $\phi$ are the amplitude, period, decay time and initial phase, respectively. The best-fit curve is shown by the thick red curve in panel (c). The period of oscillation is $P\sim$634~s, and the decay time  is $\tau\sim437$~s.

To deduce the phase speed of the wave, we require an estimate of the loop length.
In principle, STEREO images could be used to obtain loop length. 
Unfortunately, the hot loop emission was only visible in the 131 and 94~\AA\ filters, and was not visible in STEREO images.
We therefore fitted the observed loop to the circular loop model of \citet{asc2002}. 
This fits the de-projected loop with two free parameters: the height of the loop center above the solar
surface, $h_{loop}$,  and the angle between the loop plane and the vertical, $\theta_{loop}$. The method 
gives the shortest loop
compatible with the chosen points along the loop (tie points), and so the phase-speed estimate is therefore a lower bound. 
The best-fit loop with $h_{loop}=-17$\arcsec\ and $\theta_{loop}=53^\circ$, is shown in Fig.~\ref{fig2}c.
The loop length is 220\arcsec\ which implies that the wave had a phase speed $2L/P \sim510$~km~s$^{-1}$.
By varying the tie points, the shortest loop length is 200$\arcsec$ (dotted line) which gives a phase speed of 460 km s$^{-1}$.

\begin{figure*}
\centering{
\includegraphics[width=8cm]{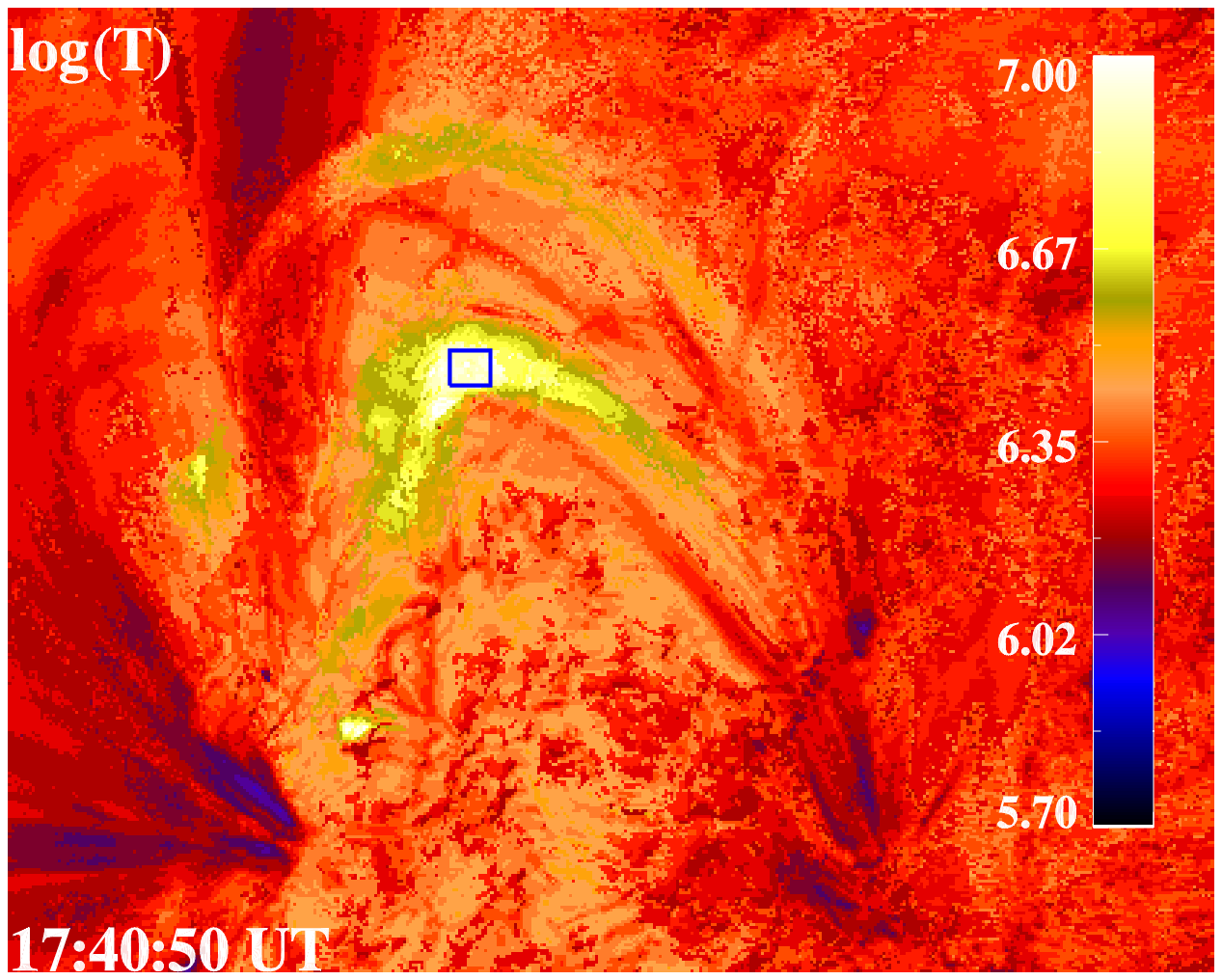}
\includegraphics[width=8cm]{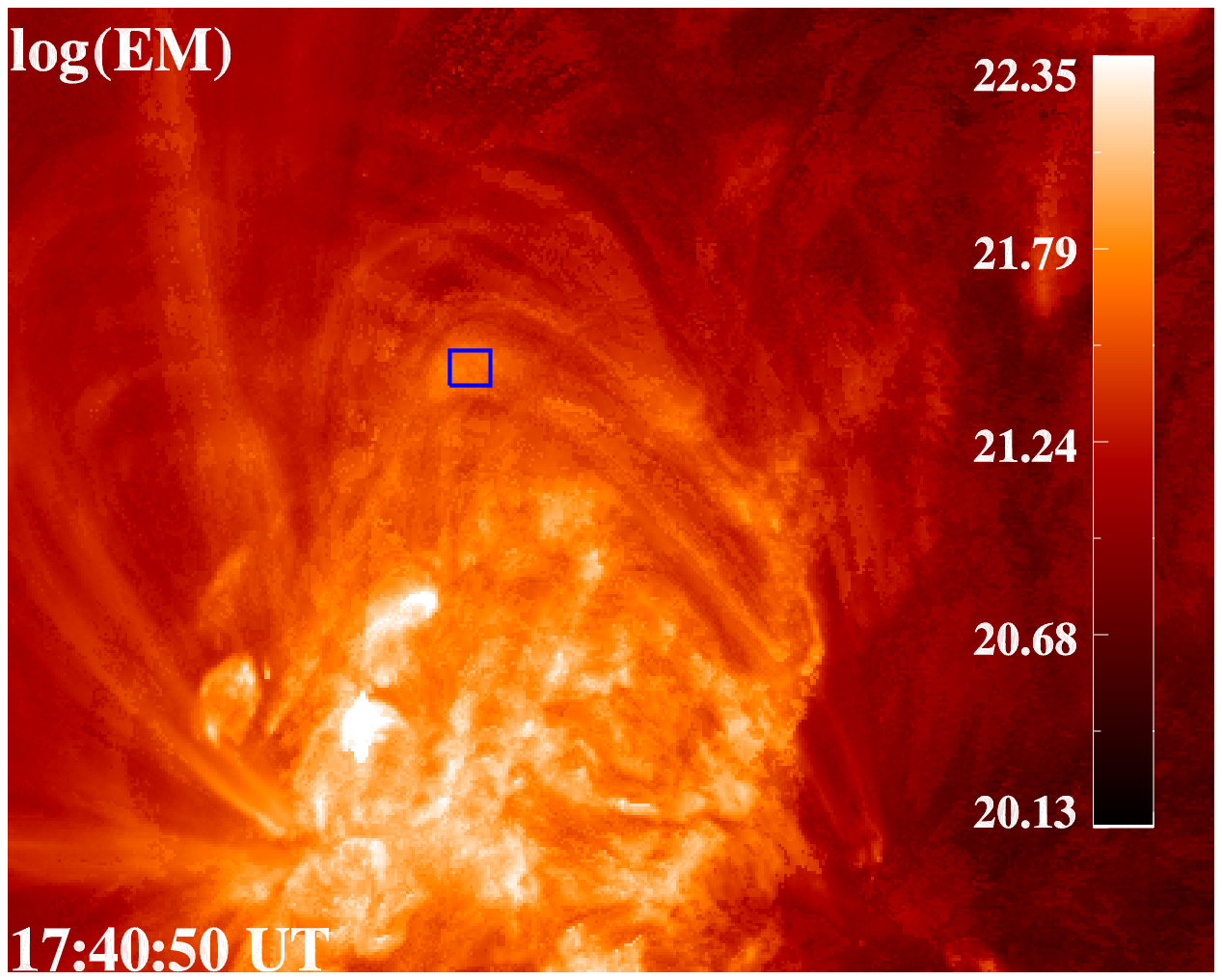}

}
\caption{Two-dimensional maps of the temperature at the peak of the DEM (log($T_p$), MK) and emission measure  (log($EM_p$), cm$^{-5}$~K$^{-1}$) derived from near-simultaneous EUV images in six AIA channels. The average peak temperature (T$_p$) and total emission measure within the box region (blue) are $\sim$8 MK and $\sim$9.96$\times$10$^{28}$ cm$^{-5}$, respectively. The size of each image is 250$\arcsec\times$200$\arcsec$.} 
\label{fig4}
\end{figure*}

To determine the temperature of the hot loop, we utilized AIA images in six EUV channels
(i.e., 94, 171, 131, 211, 335, and 193~\AA), and the SSWIDL code developed by \citet{asc2011}. In this code, the co-alignment of the AIA images from the six EUV channels is carried out by using a limb fitting method, with an accuracy of $<$1 pixel. At each position, the code fits a differential emission measure (DEM) parametrized by a single Gaussian function with three free parameters: the peak temperature emission measure ($EM_p$ in cm$^{-5}$K$^{-1}$), the temperature of the DEM peak ($T_p$), and the DEM temperature width ($\sigma_T$). The DEM peak temperatures and their emission measures across the AR are shown in Figure \ref{fig4}. At this time, the loop top, where 131~\AA\ emission was brightest, had the highest DEM peak temperature. The temperature derived in the legs represents the background active region temperature because we have not done any background subtraction. To estimate the average temperature near the loop top, we used a box region, marked in the figure, and extracted the average and maximum value of the DEM peak temperature,  $\sim$8 and 10~MK, respectively.
Using these temperatures, the sound speed within the loop was $c_s\sim$152$\sqrt{T (\mbox{MK})}\sim$430 and 480~km s$^{-1}$, respectively.

To estimate the density of the hot loop, we calculated the average values of the peak $T_p$, $EM_p$ and $\sigma_T$ in the selected region.
Using these values, we estimated the total emission measure ($EM$ in cm$^{-5}$) in the selected rectangular region
$\int \! DEM(T) \, \mathrm{d} T$.
If the depth is approximately equal to the width, $d$, of the loop \citep{cheng2012}
then the density, $n_e$, of the loop can be calculated using the relation $n_e=\sqrt{EM/d}$ (assuming the filling factor $\approx$1).
Using a total EM$\sim$9.96$\times$10$^{28}$ cm$^{-5}$, and width of the hot loop system $\sim$18$\arcsec$ (Figure \ref{fig2}),
the estimated density is $\sim$8.5$\times$10$^{9}$ cm$^{-3}$ at the top of the loop.

\subsection{Excitation Mechanism}
\begin{figure*}
\centering{
\includegraphics[width=6.5cm]{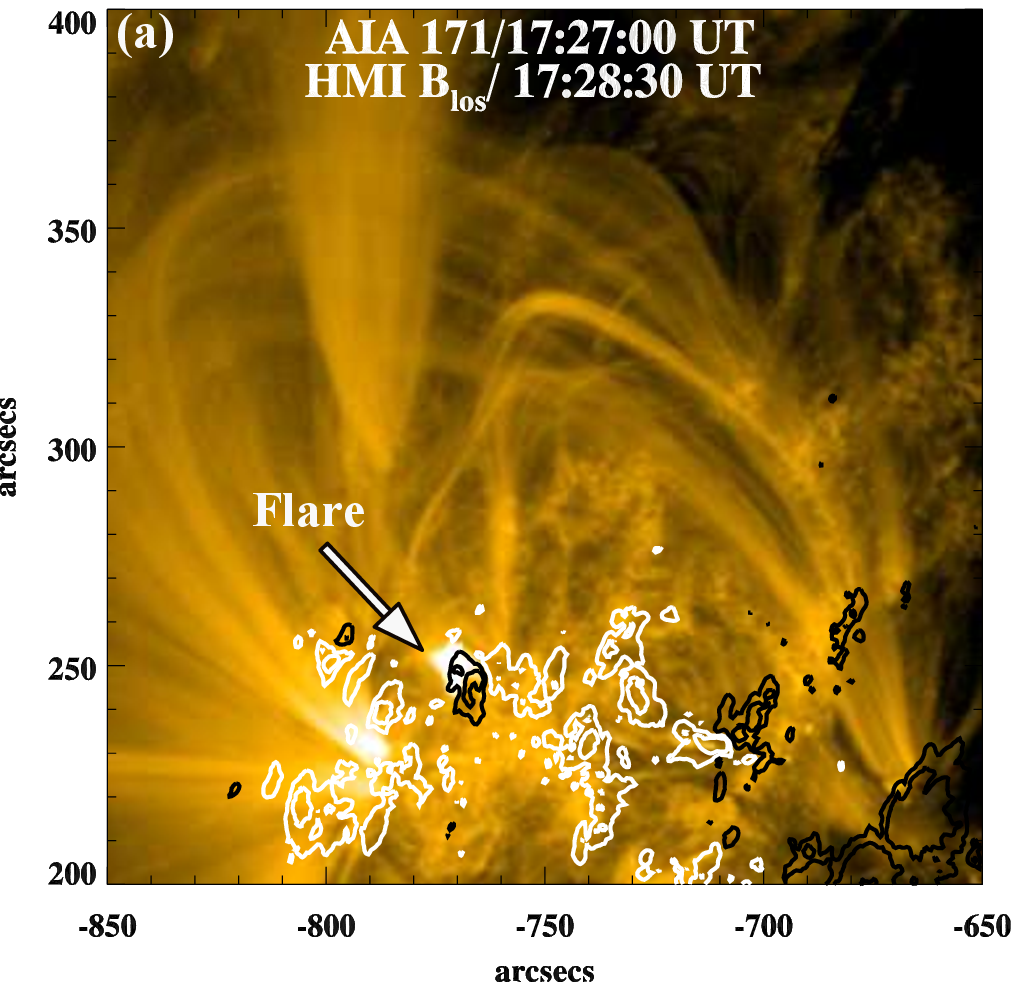}
\includegraphics[width=6.5cm]{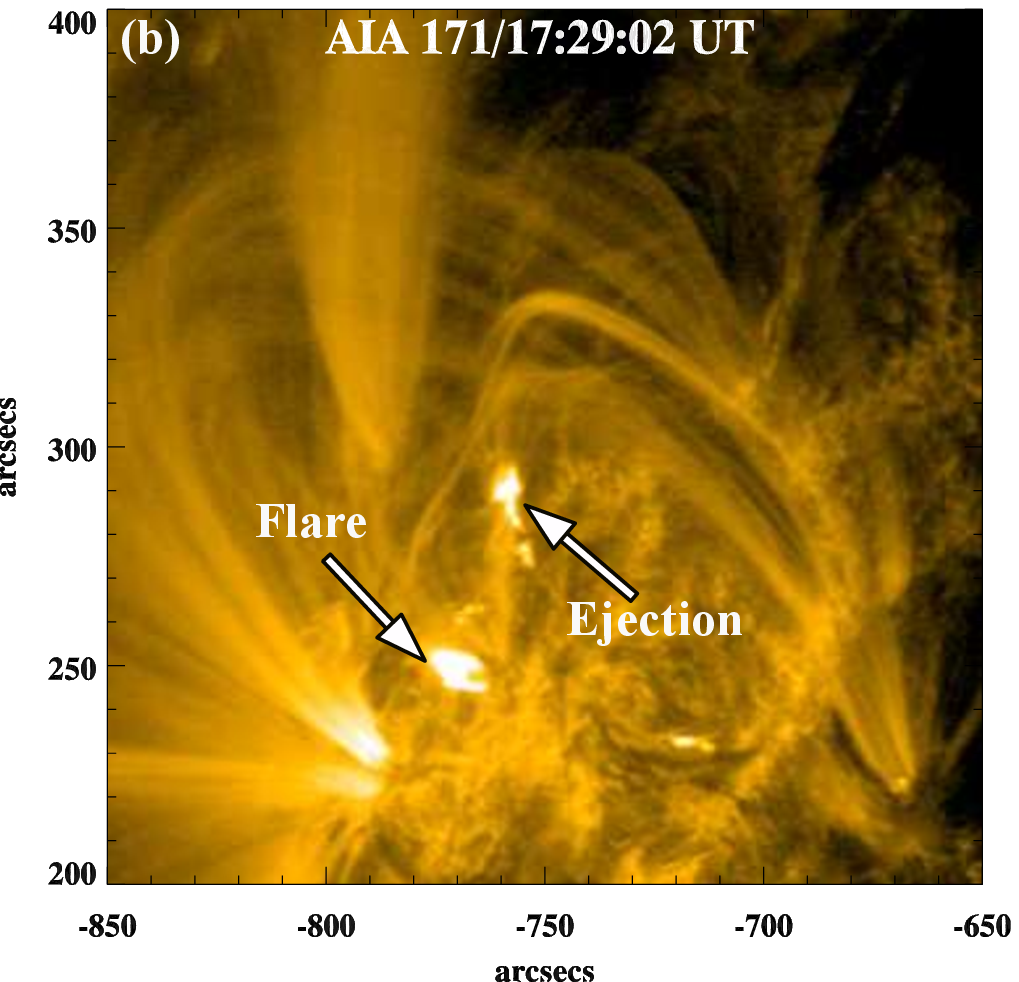}

\includegraphics[width=6.5cm]{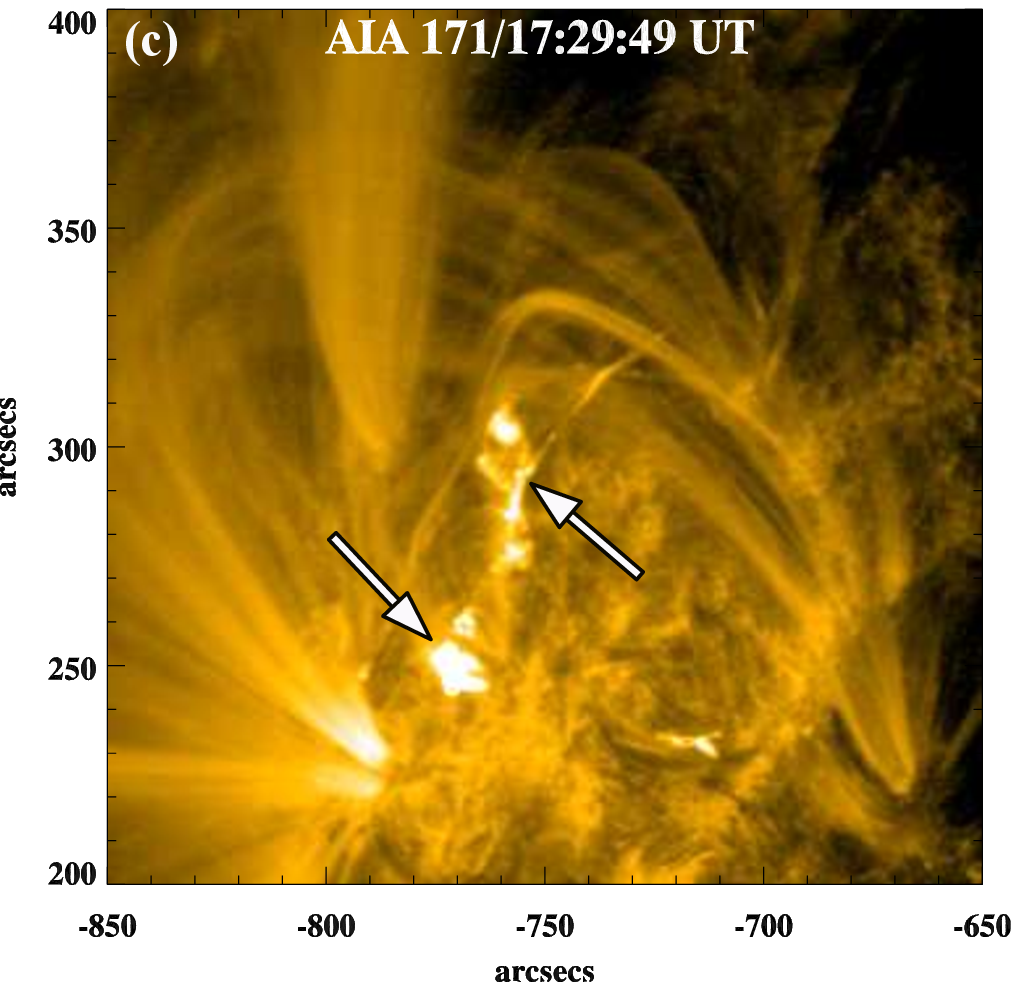}
\includegraphics[width=6.5cm]{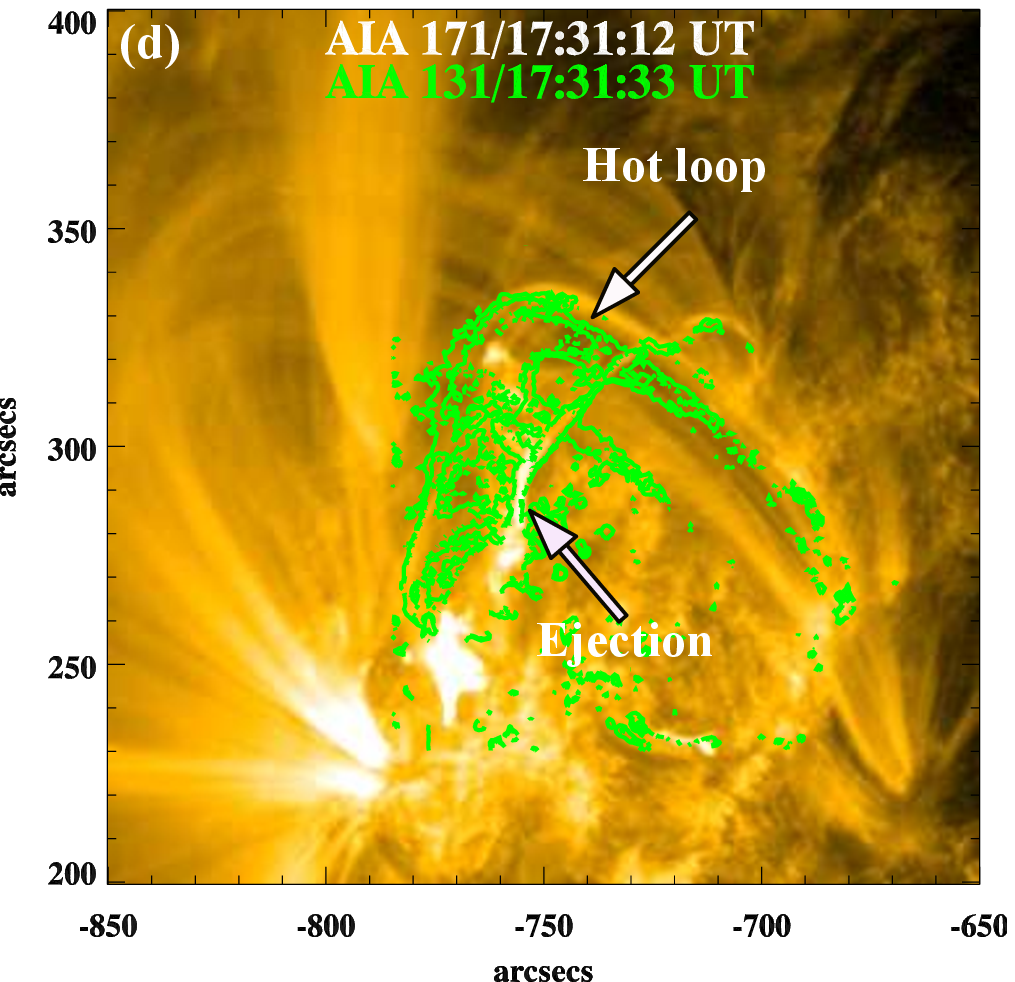}

\hspace{1.3cm}
\includegraphics[width=6.7cm]{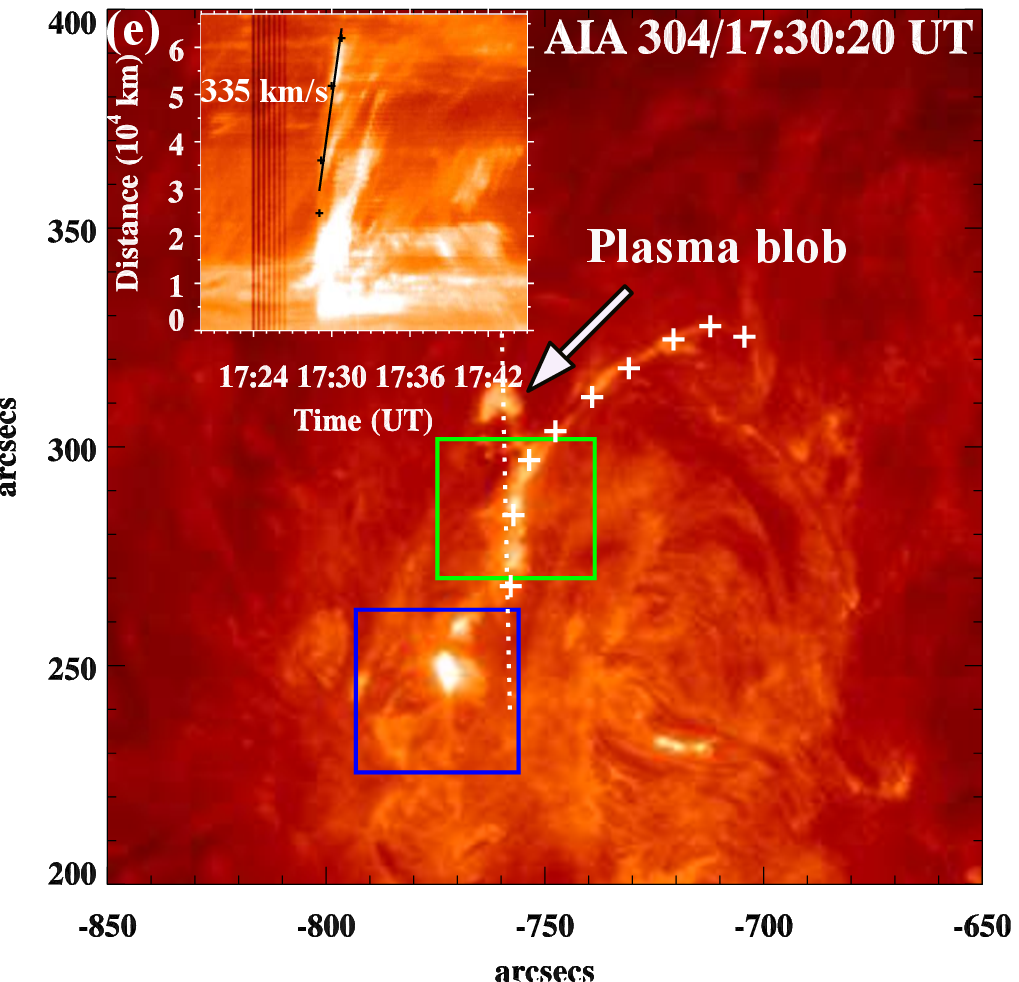}
\includegraphics[width=7.7cm]{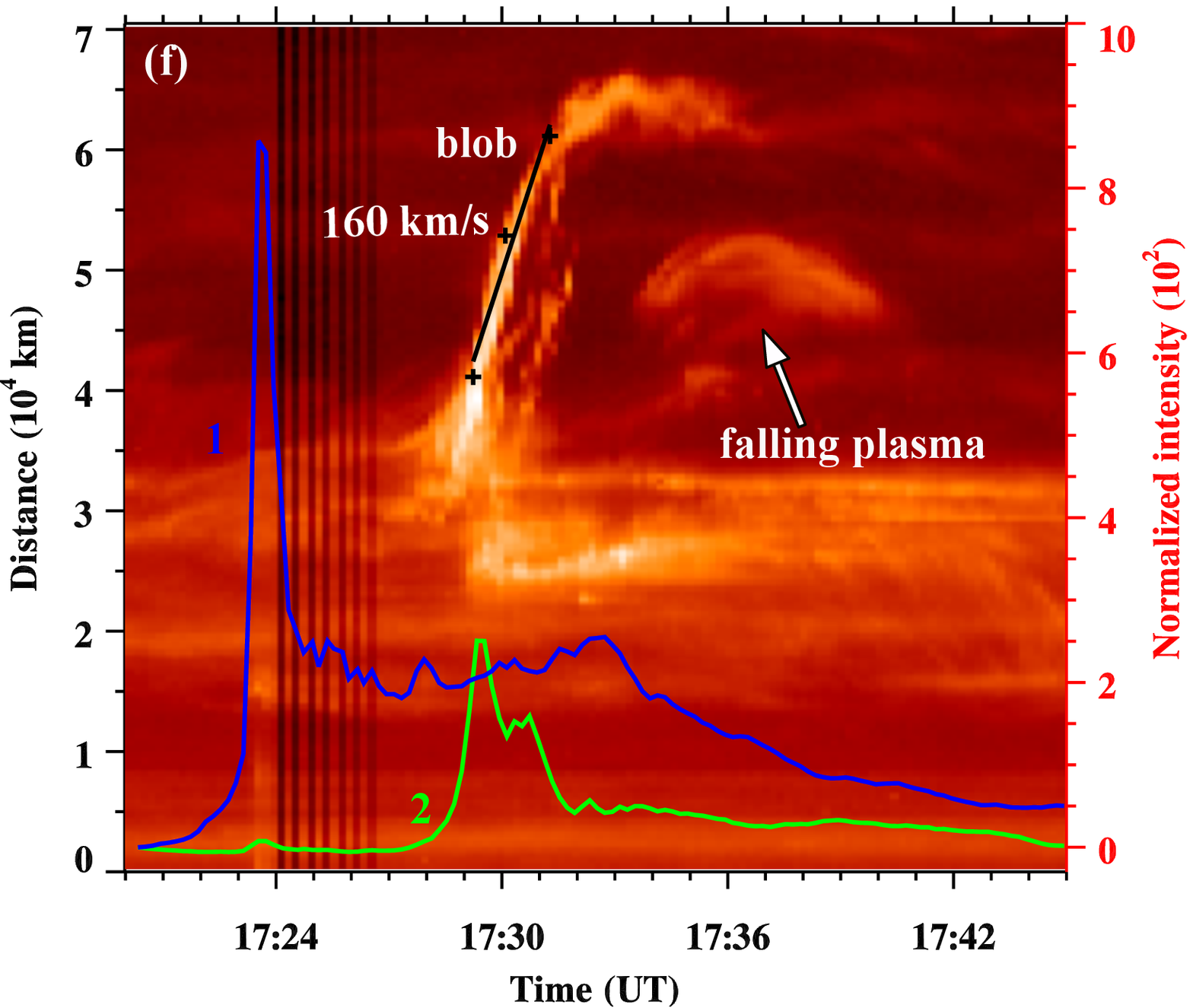}
}
\caption{(a) AIA 171 \AA~ images showing the flare site overlaid by HMI magnetogram contours. (b-c) AIA 171~\AA\ images.  (d) AIA 171~\AA\ image overlaid by 131~\AA\ (green) base-difference image contours. (e) AIA 304~\AA\ image showing the flare (blue box) and ejection of the plasma blob (along dotted line). Plasma ejection across and outside of the hot loop is marked by `+' symbols. Inset shows the space-time plot along the path (outside and across the hot loop) indicated by `+' symbols. (f) The space-time plot of the intensity distribution along the dotted line marked in panel (e) with the average AIA 304~\AA\ intensity enhancements in the sub-regions (marked in panel (e), box 1 (blue) and 2 (green)).}
\label{fig5}
\end{figure*}

To investigate the oscillation trigger, we looked at AIA 304, 171 and 131~\AA\ images. Figure \ref{fig5}(a)-(c) displays some of the selected  AIA 171~\AA\ images. Panel (a) shows the flare site (at 17:27~UT), and the magnetic configuration of the AR. The flare brightening occurred where there was a small concentration of minor, negative-polarity field. An impulsive ejection started at about 17:29:02~UT north of the flare site (panel (b) and (c)). This ejection was probably triggered by the flare. From the movies, it appeared to start from a region in the corona above the flare, and did not coincide with any strong photospheric flux concentrations. The ejection was seen in all AIA EUV channels which may be because the plasma was rapidly heated \citep{fletcher2013} or it could be due to intense emission from chromospheric lines in all the channels \citep{brosius2012}. In the 131 and 94~\AA\ movies the ejected plasma rises upwards, driving a front of hot plasma along the loop ahead of it. The front then reflected back and forth along the hot loop several times.
This is entirely consistent with the SUMER observations of two spectral components in the hot line profiles at the start of the Doppler shift oscillations.

In the cooler, 304 and 171~\AA, filter images only the plasma ejection, not the hot loop, are seen.
Panel (d) displays AIA 171 \AA~ image overlaid by AIA 131 \AA~ base difference image contours of the hot loop, which indicates that the hot loop was a separate structure, and did not overlap with  the cool 171~\AA\ loops.

At the eastern footpoint of the hot loop, a series of low-lying, hot  loops formed simultaneously with the plasma ejection.  These loops probably formed as a result of the same reconnection process that led to  the plasma ejection. We also note that brightening was seen in the 131 and 94~\AA\ images at the opposite footpoint before the arrival of the main hot plasma emission. This could indicate heating by particles accelerated at the reconnection site.

Multiple mini-plasma blobs  were ejected at the same time. They were observed in all the AIA EUV channels. The plasma ejection followed two paths: (i) along the lower edge of heated loop; (ii) across the heated loop (Figure~\ref{fig5}e).
The ejecta that went across the loop look as though they were inside the loop because they stop abruptly at the upper edge of the loop.

To estimate the speed of the plasma ejection across the loop, we looked at the 304~\AA\ intensity evolution along the ejection path (shown by a dotted line in panel (e)). The space-time plot is shown in panel (f). The plasma blob rose toward the loop-top with a speed of $\sim$160 km s$^{-1}$  which is less than the observed wave speed.
The lower part later fell back toward the solar surface.

To find the plane-of-sky speed of the ejection on the lower edge of the hot loop, we display the stack plot in panel (e) following the ejection path (`+' symbol). The speed of the blobs along this path was $\sim$335 km s$^{-1}$, which is approximately double the speed of the blob across the hot loop.

The timing of the flare, plasma ejection, and the heated loop is also illustrated by
the light curves of the average AIA 304~\AA\ intensity within the sub-regions surrounded by box 1 (blue) and box 2 (green), plotted on the stack plot in Figure~\ref{fig5}(f). In the EUV, the flare maximum is at $\sim$17:24 UT, and the plasma ejection occurred  $\sim$5 min later, at $\sim$17:29 UT.

\section{DISCUSSION AND CONCLUSION}
We report the first direct observation of an intensity oscillation along a hot loop, seen in the AIA 131 and 94~\AA\ channels.
Similar to the Doppler-shift oscillations observed by SUMER \citep{wang2002,wang2003a,wang2003b,wang2005,wang2011}, this oscillation was only seen in hot lines and had two spectral components at onset. The phase speed ($\sim$460-510~\kms) was roughly equivalent (within the errors limits of loop length and DEM temperature) to the sound speed ($\sim$430-480~\kms). The speed is consistent with a slow-mode wave which is the accepted explanation for the Doppler-shift oscillations observed by SUMER.
 In most of the SUMER events, no flare was observed prior to the waves and it was conjectured that a microflare might be the trigger.
We speculate that the oscillation was excited by a pressure pulse associated with the rapid onset of reconnection at one of the loop footpoints. 


\citet{selwa2005} numerically studied the excitation of waves in a hot ($\sim$5~MK) coronal loop by launching a pressure pulse at different positions along the loop. They found that pulses close to a footpoint of the loop excites the fundamental of the slow magnetoacoustic mode \citep{selwa2009}.  Moreover, recent MHD simulations have demonstrated that plasma flows with a subsonic speed can excite higher-speed slow-mode waves \citep{ofman2012,wang2013}.


In conclusion, we have presented a unique observational evidence of a longitudinal oscillation in a hot loop, generated by footpoint excitation.
This kind of intensity oscillation in a hot AIA loop has not been reported earlier. However, future statistical studies of the similar events using high-resolution observations from SDO/AIA and Hinode will help to understand the properties of  these waves in more detail.

\acknowledgments
We express our gratitude to the referee for his/her valuable
comments/suggestions, which improved the manuscript considerably. We thank Don Schmit for discussions. SDO is a mission for NASA's Living With a Star (LWS) program.

\bibliographystyle{apj}

\clearpage



\end{document}